\documentclass[sigconf]{acmart}

\usepackage{booktabs} 
\usepackage{float}

\acmConference[DTMBio-KMH 2018]{ACM 1st International Workshop on Knowledge Management for Healthcare}{22nd October 2018}{Lingotto, Turin, Italy}
\acmYear{2018}
\copyrightyear{2018}

\begin{document}
\title[A Relation Extraction Approach for CDS]{A Relation Extraction Approach for Clinical Decision Support}

\author{Maristella Agosti, Giorgio Maria Di Nunzio, Stefano Marchesin, Gianmaria Silvello}
\affiliation{%
  \institution{Department of Information Engineering \\ University of Padua, Italy}
  \streetaddress{Via Giovanni Gradenigo, 6/b}
  \postcode{35131}
}
\email{{maristella.agosti, giorgiomaria.dinunzio, stefano.marchesin, gianmaria.silvello}@unipd.it}

\renewcommand{\shortauthors}{M. Agosti, G. M. Di Nunzio, S. Marchesin, G. Silvello}

\begin{abstract}
In this paper, we investigate how semantic relations between concepts extracted from medical documents can be employed to improve the retrieval of medical literature. Semantic relations explicitly represent relatedness between concepts and carry high informative power that can be leveraged to improve the effectiveness of retrieval functionalities of clinical decision support systems. We present preliminary results and show how relations are able to provide a sizable increase of the precision for several topics, albeit having no impact on others. We then discuss some future directions to minimize the impact of negative results while maximizing the impact of good results.
\end{abstract}

%
%



\maketitle

\section{Motivation}
\label{sec:motivation}
The volume of medical literature published every year keeps growing at a very fast pace. 
The time required by clinicians to retrieve relevant information from such an amount of literature using standard systems is often prohibitive. Therefore, there has been a strong interest in Clinical Decision Support (CDS) systems~\cite{berner-2007} designed to produce effective and timely information that can help clinicians in the decision making process for patient care. Within this context, we focus on case-based retrieval --- i.e. given a medical case of interest, the CDS system should retrieve highly related medical literature from a large collection of medical publications. Due to severe time constraints, clinicians must take fast decisions without having the possibility to thoroughly read the literature; for this reason, case-based retrieval favors precision over recall \cite{burke_etal-2004}. 


A key characteristic of the medical literature is the large use of synonyms and context-specific expressions.
To address this term heterogeneity, Knowledge Bases (KBs) have often been exploited by Information Retrieval (IR) systems
. The current availability of medical KBs offers us the opportunity to develop techniques that better capture the semantics of medical documents, leading to the following research question: 
\begin{quote}
How can we employ the rich semantic information within medical case reports and related literature to boost retrieval performances and ease the clinical decision process?
\end{quote}

Semantic relations are a key aspect within the semantics of a document. They have been mainly used to find relevant concepts to expand a user query, but not as semantic elements to be indexed and retrieved. We hypothesize semantic relations can provide a higher semantic representation of medical cases and literature. \par

In this work, we present an initial study on the effectiveness of the use of semantic relations for the retrieval of medical literature. We define an approach comprising two methods: a \emph{rule-based} method and a \emph{learning} method. In the rule-based method, we assign a relation to a pair of concepts --- contained within the same sentence --- when it holds within a reference KB. In the learning method, we train a sentence-level relation extractor that is able to infer relation between a pair of concepts given the sentence's context. 

We evaluated our approach by using the publicly shared OH\-SU\-MED collection \cite{hersh_etal-1994}. OHSUMED provides rather short queries which represent a hard task for our approach, since limited information --- e.g. concepts and relations --- can be extracted from them. Testing with OHSUMED allow us to assess the potential and limitations of the approach. 
The remainder of the paper is organized as follows: Section \ref{sec:related_work} presents the background and related work, Section \ref{sec:approaches} describes the proposed approach, Section \ref{sec:experiments} presents experiments and results and Section \ref{sec:discussion} draws some conclusions and outlines future work. 

\section{Related Work}
\label{sec:related_work}
\par

Concept-based IR aims at making use of external sources (like thesauri and ontologies) to provide additional knowledge and context that may not be explicit in a document collection and users' queries. 
Concept-based methods can be categorized in two types: (i) methods that use concepts in both indexing and retrieval stages \cite{egozi_etal-2011}, and (ii) methods that apply concept analysis in one specific stage, such as concept-based query expansion \cite{grootjen_van-der-weide-2006}. The approach we adopt extends the use of concepts to relations and uses them in all stages of retrieval. This is more challenging, but it allows for a finer semantic representation of documents and queries. \par

In the biomedical domain --- where there are authoritative and curated ontologies --- concept-based approaches demonstrate consistent improvements over classic keyword-based systems. 
In \cite{koopman_etal-2012}, 
`is-a' relationships between concepts are used to weight documents containing concepts subsumed by the query's concepts.
\cite{limsopatham_etal-2013} proposes a method to represent medical records and queries by focusing only on medical concepts essential for the information need of a medical search task. In \cite{limsopatham_etal-2013a}, queries are expanded by inferring additional conceptual relationships from domain-specific resources as well as by extracting informative concepts from the top-ranked medical records.\par 

The field of Biomedical Information Extraction (BioIE) 
is highly relevant for CDS. 
\cite{liu_etal-2016} reviews the recent advances in learning-based approaches for BioIE tasks. BioIE tasks comprise entity linking \cite{zheng_etal-2015}, event identification \cite{ananiadou_etal-2010} and relation extraction \cite{uzuner_etal-2011,wang_fan-2014}. 
Being targeted to CDS --- i.e. voted to the extraction of key relations that can facilitate clinical decision making --- our problem setup is fundamentally different from the conventional biomedical setups. Most of state-of-the-art biomedical relation extraction techniques are developed for specific relations, like protein-protein interactions, gene-disease interactions and so on --- 
which cover only a fraction of the biomedical domain. 

Regarding relations in IR, \cite{voskarides_etal-2017} study the problem of finding human readable descriptions of a given relationship in a knowledge graph. \cite{schuhmacher_etal-2016} apply supervised relation extraction to documents that are relevant for an information need Q and study how many of the extracted relations are indeed relevant for Q. \cite{kadry_dietz-2017} explores current state of the art in unsupervised relation extraction (OpenIE) for the task of finding support passages to complement an entity ranking with human-readable explanations of how those retrieved entities are connected to the information need. Conversely, our approach applies supervised relation extraction to extract semantic relations that are used in all stages of retrieval. Hence, relations play a pivotal role in the actual retrieval of documents.

\section{Methodology}
\label{sec:approaches}
We present a new approach that uses semantic relations for case-based retrieval. The methodology is composed of the information extraction step (Subsection \ref{subsec:IE}) and the information retrieval step (Subsection \ref{subsec:IR}).

\subsection{Information Extraction}
\label{subsec:IE}
The information extraction step is divided into an entity linking component and a relation extraction component. \par 

The entity linking component extracts entity mentions within the text and links them to a reference KB; this reduces the high number of synonyms, abbreviations and context specific expressions that are present in the medical literature. For entity linking we adopt MetaMap,\footnote{\url{https://metamap.nlm.nih.gov/}} the most authoritative tool to detect medical entity mentions in free-text. MetaMap analyses biomedical free-text and identifies concepts belonging to the Unified Medical Language System (UMLS), associating each mention with a number of concepts from the UMLS Metathesaurus\footnote{\url{https://www.nlm.nih.gov/research/umls/knowledge_sources/metathesaurus/}} --- which comprises more than 3 million distinct concepts. Within UMLS, a substantial understanding of the medical domain is included, comprising medical concepts, relations, definitions and so on. \par

The relation extraction component detects semantic relation between pairs of concepts within a sentence. 
To be consistent with concepts extracted with MetaMap, we consider semantic relations from UMLS Metathesaurus as well. Furthermore, since our task requires a high coverage of the medical domain, considering UMLS Metathesaurus relations --- which are coarse-grained relationships that span to a high number of concepts ---  allows us to increase the recall of extracted relations. \par 

We define two methods for the extraction of relations from documents and queries: a rule-based method and a learning method. \textbf{Rule-based:} a relation is assigned to a pair of concepts if it relates them within UMLS. We assume that a UMLS relation between two concepts always occurs, even when it is not explicitly mentioned in the sentence containing the two concepts. \\
\noindent \textbf{Learning:} we train a distantly supervised \cite{mintz_etal-2009} sentence-level Bidirectional Long Short-Term Memory (BiLSTM) neural network to detect if a relation exists between two concepts based on the context of the sentence. The network architecture is composed of an input (word embedding) layer of concatenated word features and positional features. Words are first converted into pre-trained word embeddings trained on 26 million abstracts and citations in PubMed --- released by \cite{moen_etal-2013}. Then these word features are concatenated with two sets of positional features --- to explicitly account for the pairs of words to which we expect to assign relations \cite{zeng_etal-2014}. We apply a max-pooling layer right after the bidirectional recurrent layer and before the output layer --- in order to combine segment-level features that, although not very strong in representing the entire sentence, represent local patterns well \cite{zhang_wang-2015}. In this way, we try to overcome the tendency of recurrent connections to forget long-term information too quickly, leading the supervision at the end of the sentence to be hardly propagated to early steps in model training (due to gradient vanishing \cite{bengio_etal-1994}).

\begin{figure*}[t]
\center
\includegraphics[width=\textwidth]{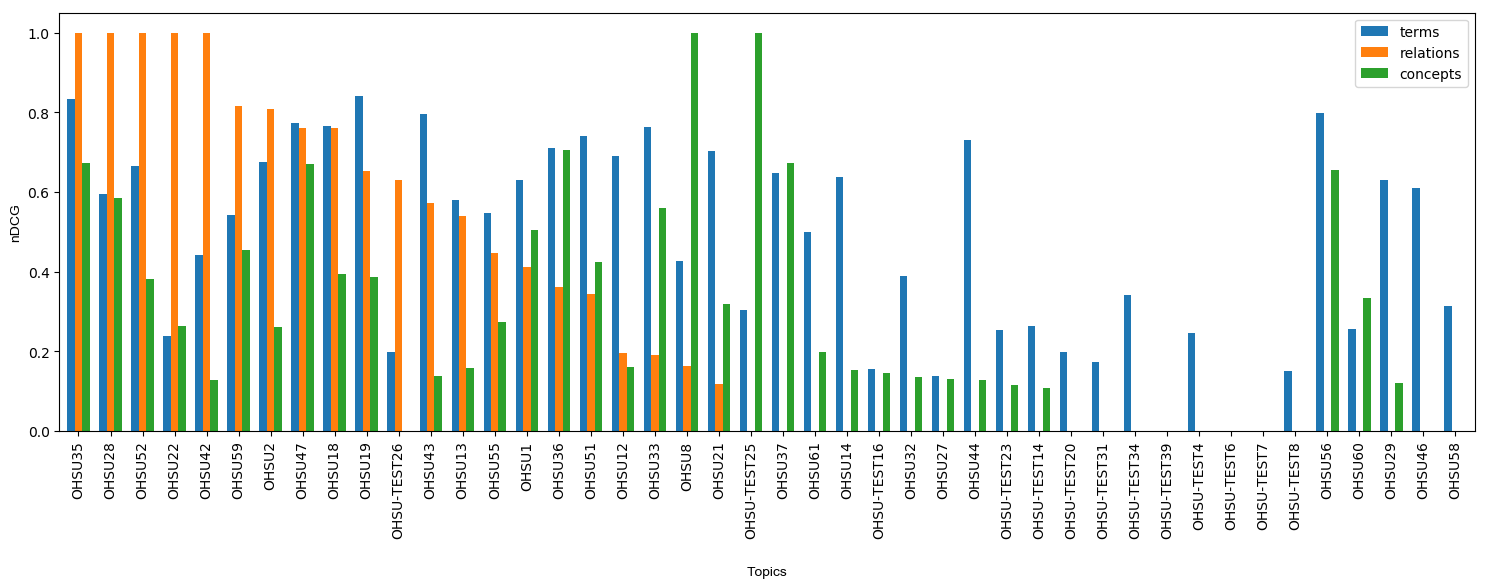}
\caption{nDCG values for topics containing relations. OHSU\{56,60,29,46,58\} returned NA values for the BoR representation. Queries are sorted in descending order first by BoR (relations) nDCG values, then by BoC (concepts) nDCG values.}
\label{fig:ndcg_values}
\end{figure*}

\subsection{Information Retrieval}
\label{subsec:IR}
Documents are indexed by considering all terms as in the Bag-of-Words (BoW) representation. We extend the BoW representation to both concepts (BoC) and relations (BoR) by considering for the indexing all the extracted concepts and relations respectively. The ranking is obtained using Okapi BM25 ranking formula \cite{robertson_walker-1999}. \par

Since relations are extracted at sentence level, we also index passages --- i.e. groups of consecutive sentences --- by considering all the relations occurring within each group of sentences (passage-level BoR). Relevant passages should contain a higher number of relations related to the information need when compared to non relevant passages --- being more similar in their semantic contents to the query. Therefore, documents that contain more relevant passages can be considered more relevant for the query. \par

We define a weighting scheme such that a document score is computed as the weighted sum of its passages scores, where scores are computed using BM25 as above. The passage-level weighting scheme is as follows: \par 

\begin{equation}
score(q,d) = \sum_{p \in d} \frac{|R_p \cap R_q|}{|R_q|} BM25(p, q)
\end{equation}

\noindent where $d$ is the document, $q$ is the query, $p$ is a passage belonging to document $d$, $R_q$ is the set of relations extracted from query $q$ and $R_p$ is the set of relations extracted from passage $p$. 

\section{Experiments and Results}
\label{sec:experiments}
We employed the OHSUMED test collection which contains 348,566 references from the on-line medical information database MEDLINE, consisting of titles and/or abstracts from 270 medical journals over a five-year period (1987-1991). The available fields are: title, abstract, MeSH indexing terms, author, source, and publication type. There are 106 queries in the collection. Each query is composed of two sentences: title + description. Title is the brief summary of the medical case at hand, description is the information need required to answer a specific question for the case.

\emph{\textbf{Experimental Setup:}}
We performed two experiments: i) one using the rule-based method to extract relations out of documents and queries; ii) the other using the learning method to extract relations out of documents and queries. We compared the results obtained applying BM25 to the three representations (i.e. BoW, BoC and BoR) and we evaluated the results using the nDCG measure. 

\emph{\textbf{Results:}}
i) The rule-based method was able to extract relations from a subset of 44 queries. Therefore, to investigate the effectiveness of relations, we restrict the experiments to this subset only --- since the remaining queries lead to no results when considering relations. Of these 44 queries, only 39 have relations matching with some documents. Regarding the relations, we obtained the best results with the passage-level approach. We set the passage length to 2, in order to be compliant with queries' length. Documents' score was computed using the formula shown above (1). 
The nDCG results on these 39 queries are variable --- ranging from 0 (18 cases) to 1 (5 cases), as can be seen in Figure \ref{fig:ndcg_values}. 

Such a variance gives us some hints about the informative power of relations. When properly extracted, relations can be highly effective, indeed, we compared the average nDCG values of concepts and relations on only those topics where relations give a result different than 0 and we found a statistically significant average improvement of 20\%. A \emph{t-test} was performed to validate the improvement. Regarding the comparison between relations and terms, the behavior of relations is similar to the one of terms (baseline approach), and there is no statistically significant difference between the two. \par

ii) The learning method was able to extract relations from a subset of 25 queries. Of these 25 queries, only 12 have relations matching with some documents. The results on these 12 queries are comparable to those presented for the rule-base method, with nDCG values ranging from 0 (in 7 cases) to 1 (in 1 case). The reason for this is two-fold: (a) the shortness of queries that limits the relations that can be extracted; and, (b) the highly different syntactic structure of queries if compared to the sentences within the medical abstracts leading to a mismatch between the query-relations  and abstract-relations.

\vspace{-0.1cm}

\section{Discussion}
\label{sec:discussion}
In this work, we proposed and evaluated the effectiveness of semantic relations as basic constituents for a CDS system. We defined two methods for extracting relations from queries and documents: a rule-based method and a learning method. 
We found that relations --- when pertinent to the initial information need --- are highly valuable, outperforming concepts. The challenge lies in how to limit those cases where relations provide no relevant results for the information need. To this end, considering collections where queries present a long and narrative structure (e.g. TREC CDS tracks) might be a possible direction to balance such issue. Furthermore, defining more IR-oriented relation extraction algorithms that are capable of overcoming the high precision-low recall nature of state-of-the-art methods is a direction we will investigate. 
%


\begin{acks}
Supported by the CDC-STARS project of the University of Padua.

\end{acks}
\twocolumn
\newpage
\bibliographystyle{ACM-Reference-Format}
\bibliography{Marchesin}

\end{document}